\documentclass[conference]{IEEEtran}
\usepackage{xspace,shortcuts}
\usepackage{comment}
\usepackage{graphicx}
\usepackage{algorithm}
\usepackage{algorithmic}
\usepackage{amsmath}
\usepackage{amssymb}
\usepackage{float}
\usepackage{multirow}
\usepackage{courier}		
\usepackage{url}
\usepackage{listings}
\usepackage{xcolor}
\usepackage{subcaption}

\usepackage{booktabs} 

\usepackage{color,soul}
\usepackage{listings}

\definecolor{backcolour}{rgb}{0.95,0.95,0.92}

\begin{document}
\title{Build and Execution Environment (BEE):
an Encapsulated Environment Enabling
HPC Applications Running Everywhere}

\author{Jieyang Chen$^{*}$, Qiang Guan$^{\ddag}$, Xin Liang$^{*}$, Paul  Bryant$^{\ddag}$ Patricia Grubel$^{\dag}$, Allen McPherson$^{\dag}$,\\
  Li-Ta Lo$^{\dag}$,   
Timothy Randles$^{\dag}$, Zizhong Chen$^{*}$, and James Paul Ahrens$^{\dag}$ \\
$^{*}$University of California, Riverside 
$^{\ddagger}$Kent State University 
$^{\dag}$Los Alamos National Laboratory \\
\{jchen098, xlian007, chen\}@cs.ucr.edu, 
\{qguan, pbryant1\}@kent.edu, \\
\{ pagrubel, mcpherson, ollie, trandles, ahrens\}@lanl.gov
}


\maketitle

\linespread{0.97}%
\selectfont

\begin{abstract}
Variations in High Performance Computing (HPC) system software configurations mean that applications are typically configured and built for specific HPC environments. Building applications can require a significant investment of time and effort for application users and requires application users to have additional technical knowledge. Linux container technologies such as Docker and Charliecloud bring great benefits to the application development, build and deployment processes. While cloud platforms already widely support containers, HPC systems still have non-uniform support of container technologies. In this work, we propose a unified runtime framework -- Build and Execution Environment (\texttt{BEE}) across both HPC and cloud platforms that allows users to run their containerized HPC applications across all supported platforms without modification. We design four \texttt{BEE backends} for four different classes of HPC or cloud platform so that together they cover the majority of mainstream computing platforms for HPC users. Evaluations show that \texttt{BEE} provides an easy-to-use unified user interface, execution environment, and comparable performance.
\end{abstract}



\begin{IEEEkeywords}
high performance computing; cloud computing; container.
\end{IEEEkeywords}

\section{Introduction}
  \label{sec:introduction}

High Performance Computing (HPC) systems have become critical infrastructures for science and industry. For example, domain experts use HPC systems to run large-scale physical simulations, big data analysis, multi-layer artificial neural networks, molecular dynamics experiments, and DNA sequencing.

As different HPC systems typically have customized software environments, HPC users must often configure and build their application for each specific machine, which is time consuming and can become a bottleneck to productivity. Moreover, in some cases, years- or even decades-old legacy applications still serve as key components in the process of scientific or industry research.  These legacy applications may have no active support and often require specific deprecated versions of libraries and/or hardware in order to make them run correctly. It may be hard to find or build library implementations that meet the requirement of legacy applications while remaining compatible with current HPC systems. These requirements impose great challenges for HPC users.  


Due to availability of shared computing environment, users may need to frequently switch between computing systems. This can be among different HPC platforms or even cloud platforms. Differences in software environments may impede users when choosing a platform that has different software/hardware configurations. A consistent execution environment is also important for HPC application developers. Consistency between developing, testing, and production environments can greatly save developers' time on fixing compatibility issues, that can significantly accelerate the development process.

\begin{figure*}[h!]
    \centering
    \includegraphics[width=0.8\textwidth]{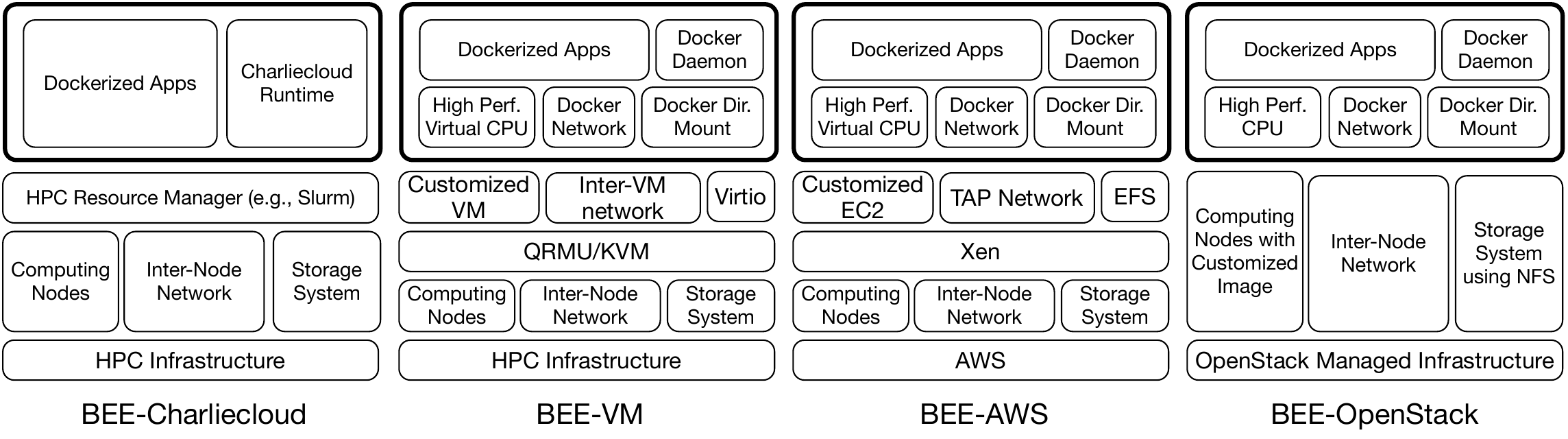}
    \caption{BEE Backends}
    \label{bee-backend}
\end{figure*}

Virtualized environments, and particularly virtual machines (VMs), have been thoroughly investigated for HPC systems to provide more consistent, isolated, and secure environments \cite{vallee2008system, reuther2012hpc}. For example, Huang, et al. \cite{huang2006case} built a virtualized HPC environment using Xen-based VMs. By leveraging high performance I/O passthrough \cite{liu2006high}, VMs can achieve near-native performance when running HPC benchmarks. In \cite{zhang2016slurm}, Zhang, et al. illustrated that current resource managers in HPC systems cannot well supervise VMs and associated critical resources, so they proposed Slurm-V, that extends Slurm with virtualization-oriented capabilities. Huang, et al. and Tikotekar, et al. \cite{gugnani2016performance, tikotekar2008analysis} characterized the performance  of running multiple kinds of applications on virtualized HPC clusters. In \cite{huang2007virtual}, Huang, et al. proposed Inter-VM Communication (IVC), a VM-aware communication library to support efficient shared memory communication among computing processes on the same physical host and then they built an MPI library that was IVC enabled.

However, managing application specific VM images is not trivial, given that VM images need to contain files of the operating system, dependent libraries/packages, user applications, and input/output data, that may take several gigabytes of disk space. Migrating images between HPC systems or distributing images among compute nodes can consume a lot of time. On the other hand, Linux containers are more lightweight. Many implementations of Linux containers have been proposed e.g., Docker \cite{Docker}, Charliecloud \cite{priedhorsky2016charliecloud}, Shifter \cite{jacobsen2015contain}, and Singularity \cite{kurtzer_2016_60736}. They provide consistent execution environments for development, build, and deployment. By using Linux containers, developers only need to build their application once in the container on their local machine, and then the application can run on any supported machine. For doing that, one only needs to transfer the container image to the target execution machine. The container image usually only contains minimal operating system composition, application-dependent libraries/packages, and user applications, with much less space requirement than VM images \cite{boettiger2015introduction}. Also, since  Linux containers allow the guest application to share much of the host operating system including the Linux kernel, the performance penalty is small \cite{merkel2014docker, ruan2016performance}.

It can be of great benefit to bring the advantages of 
Linux containers, as realized in the cloud, to HPC users; however, HPC systems have support of Linux containers in a non-uniformed fashion. Some systems can natively support widely used Docker containers; Some systems run incompatible older Linux kernels that make them unable to support Linux containers \cite{harji2013our}. Some systems cannot support Docker containers but can support other Linux container implementations e.g., Charliecloud \cite{priedhorsky2016charliecloud}, Shifter \cite{jacobsen2015contain} and Singularity \cite{kurtzer_2016_60736}. This is challenging for HPC users who want to take the benefits of Linux containers and who already have their HPC applications running in Linux containers on one system and need to migrate to another one, since different Linux containers/runtimes are not mutually compatible with each other.  Even on an HPC/cloud system where a certain kind of Linux container is supported, running HPC applications still needs complicated configuration, since Linux containers, by their nature, are built for isolation, that is contrary to the sharing fashion adopted by HPC applications.




In this work, we design a unified build and execution environment that overcomes the challenges when running containerized applications in HPC systems.  We call it Build and Execution Environment (\texttt{BEE}).

\begin{enumerate}
\item \textbf{Docker image support:} Among all kinds of Linux container implementations, Docker is one of the most widely used implementations in the HPC community. With the support of Docker image, current HPC users can deploy their Dockerized application using \texttt{BEE} with no modification.
\item \textbf{Reproducibility:}
\texttt{BEE} aims to build up similar HPC-friendly execution environments across different platforms, allowing Dockerized applications to behave consistently. This is accomplished through a series of specially design modules -- \texttt{BEE backends}. Different \texttt{BEE backends} target different classes of platform, but they can build up execution environments with similar hardware configuration and software stack.
\item \textbf{Multiple platforms support:}
We design four \texttt{BEE backends} that support four different classes of systems. For HPC systems, instead of using Docker container runtime, we choose to use Charliecloud, a Linux container implementation with much less usage requirement than Docker and its runtime supports running Docker images. Charliecloud only needs execution systems to have Linux user namespace enabled, and it is usually enabled by default on many current and new HPC systems. Using Charliecloud, we build our first \texttt{BEE backend} for HPC systems -- \texttt{BEE-Charliecloud}. For older HPC systems that do not have Linux user namespace support, \texttt{BEE} provides another \texttt{BEE backend} -- \texttt{BEE-VM}. It can run Dockerized HPC applications through Docker runtime via a specialized VM. In addition to HPC systems, we also design two \texttt{BEE backends} to support running HPC applications on cloud systems. These allow users to run HPC applications when cloud platforms are preferable computing resources for users. Specifically, \texttt{BEE} provides \texttt{BEE-AWS} that allows Dockerized HPC applications to run on Amazon Web Services (AWS) platform and \texttt{BEE-OpenStack} for OpenStack-based HPC or cloud infrastructures. 
\item \textbf{End-to-end automation:}
\texttt{BEE} provides end-to-end automation that hides all launching details from users. Users only need to provide a \texttt{BEE} task description file (\texttt{beefile}), Docker images, and run scripts in order to launch a task on \texttt{BEE}. \texttt{BEE} will handle all the complications including: connecting to the remote platform, setting up suitable computing environment, configuring network and storage, and launching applications.
\item \textbf{Flexibility:}
\texttt{BEE} provides a unified user interface, so together with \texttt{BEE backends}, \texttt{BEE} users have the flexibility to switch between different platforms, as their needs require, with no modification to their applications and minimum modification to the task description file.
\end{enumerate}

\begin{figure*}[t]
    \centering
    \includegraphics[width=0.8\textwidth]{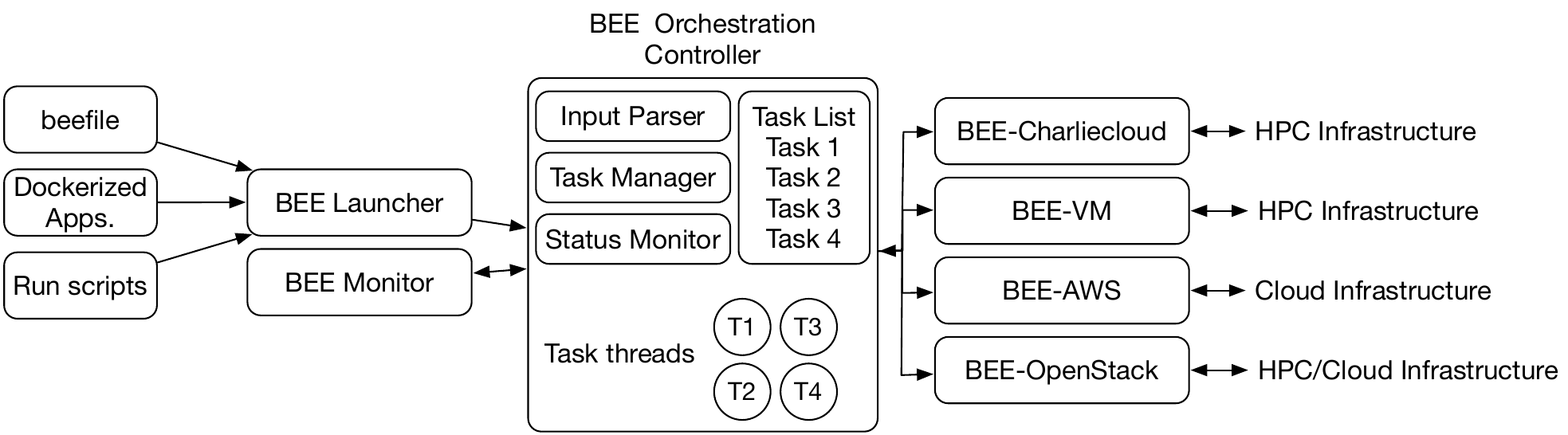}
    \caption{BEE Framework}
    \label{bee-framework}
\end{figure*}

The rest of this paper is organized as follows:  
We discuss the BEE framework design in section \ref{bee-framework-section}. The designs of four \texttt{BEE backends} are discussed in section \ref{bee-charliecloud-section} - \ref{bee-openstack-section}. Performance is evaluated in section \ref{evaluation-section}. We showcase VPIC, a real HPC application running in \texttt{BEE} in section \ref{case-study-section}. In section \ref{related-work-section}, we discuss related works and how \texttt{BEE} is unique. Finally, we make our conclusion in section \ref{conclusions-section}.

\section{BEE design overview}
\label{bee-framework-section}
The goal of \texttt{BEE} is to enable unified experience for users when launching HPC applications across different platforms. This is manifested in two ways: (1) A unified user interface that only requires minimum configuration to launch applications on a platform or switch between different platforms; (2) Similar execution environments for applications on different platforms that enable consistent application execution behavior.

\subsection{User interface}

The unified user interface consists of three inputs when launching an application using \texttt{BEE}:
\begin{enumerate}
\item A Docker image containing the application;
\item Run scripts;
\item Task description file -- \texttt{beefile};

\end{enumerate}
The first input is the Docker image containing the application. Then, to run the application when it is deployed on a platform, users need to provide run scripts. Depending on the application, the scripts can be in many forms e.g., Shell scripts, Python scripts, etc. Both the Docker image and run scripts do not need to be modified when switching between different execution platforms. Finally, to hide most complications and still provide enough capability of customization, we propose to use a simple JSON-format task description file (\texttt{beefile}) to handle all the communications between users and \texttt{BEE}. \textbf{Listing 1} shows the template of a \texttt{beefile}. Users need to select the suitable execution platform (line 3), provide general sequential run scripts (line 5 - 9) and MPI parallel run scripts (line 10 - 14), information about Dockerized  application (line 16 - 20), and finally provide necessary platform-specific information (line 21 - 26) e.g., node list, credential information. Note different Docker containers, by default, do not share a file system at runtime, but many HPC application processes running on different nodes need to have a shared directory to share data, so \texttt{BEE} needs users to specify a directory inside the container (line 16) that will be mounted to a shared host filesystem at runtime. When switching between platforms, users only need to make modifications to the \texttt{beefile}.

\lstset{numbers=left,
xleftmargin=1.5em,
frame=single,
framexleftmargin=2em}

\lstset{language=Java}
\begin{lstlisting}[ float, escapechar=!,
				   caption= Template of \texttt{beefile}]
 "task_conf": {
  "task_name": <task name>,
  "exec_target": bee_cc|bee_vm|bee_aws|bee_os,
  "general_run": [
  	{ "script": <script 1> },
   { "script": <script 2> }, ...
  ],
  "mpi_run": [
  	{ "script": <script 1> },
   { "script": <script 2> }, ...
  ]
 },
 "docker_conf": {
  "docker_img_tag": <docker image>,
  "docker_username": <username>,
  "docker_shared_dir": <dir>
 },
 "exec_env_conf": {
  "bee_cc": {...} or
  "bee_vm": {...} or
  "bee_aws": {...} or
  "bee_os": {...} 
 }       

\end{lstlisting}

\subsection{Execution environment}
To build similar execution environments on different platforms that can lead to consistent application execution behavior, \texttt{BEE backends} are proposed. A \texttt{BEE backend} is a module in \texttt{BEE} framework that can automatically build up an HPC-friendly execution environment on a specific class of platforms. We design different \texttt{BEE backends} for different classes of platforms that can build up similar execution environment. As for now, there are four \texttt{BEE backends}: \texttt{BEE-Charliecloud}, \texttt{BEE-VM}, \texttt{BEE-AWS}, and \texttt{BEE-OpenStack}. These are built aiming at four different classes of platforms: \texttt{BEE-Charliecloud} supports current and later HPC systems that have Linux user namespace support; \texttt{BEE-VM} supports older HPC systems; \texttt{BEE-AWS} supports AWS cloud platform; \texttt{BEE-OpenStack} supports OpenStack-based HPC or cloud platforms.

\textbf{Fig. \ref{bee-backend}} shows the hardware and software stack of the four \texttt{BEE backends}. The bold rectangle on the top row of each \texttt{BEE backend} indicates the execution environment that each \texttt{BEE backend} provides. Below that are different technologies that are used to enable consistent execution environment. Further below are the three components that are most critical to HPC applications: \textit{computing}, \textit{network}, and \textit{storage}. Each \texttt{BEE backend} is designed to provide similar computing, network, and storage environment for the runtime environment that runs Dockerized HPC applications. This makes sure that applications do not need to be modified to run on another \texttt{BEE backend} and preserving the same execution behavior. We will discuss the design details of each \texttt{BEE backend} in the following sections.

\subsection{Components in \texttt{BEE} framework}
Finally, we introduce other components in the \texttt{BEE} framework that help connect the user interface with  \texttt{BEE backends}. \textbf{Fig. \ref{bee-framework}} shows the framework of \texttt{BEE}. The \texttt{BEE Launcher} is the \texttt{BEE frontend} for users to launch a task on \texttt{BEE}-supported platforms, using the three inputs mentioned earlier. The core part of \texttt{BEE} is the \texttt{BEE Orchestration Controller}, that is responsible for connecting the \texttt{BEE frontend} and the four \texttt{BEE backends}. It takes the task launching requests from \texttt{BEE Launcher} and initiates task launching process through \texttt{BEE backends}. \texttt{BEE Orchestration Controller} uses different threads to manage different tasks at the same time. It also keeps track of the status of each launching process.

\section{BEE-Charliecloud}
\label{bee-charliecloud-section}
Charliecloud \cite{priedhorsky2016charliecloud} is a Linux container runtime based on Linux user namespaces. It offers all the necessary runtime functionalities for HPC applications. It has three major benefits that make it the ideal runtime for \texttt{BEE} on HPC systems: First, it takes standard Docker images as input via the built-in image flattener, so it helps maintain the same user interface as the other three \texttt{BEE backends}. Second, Linux user namespaces only bring minor performance overhead, full exposure of host hardware resources, and no requirement of privileged operations or daemons. Third, Linux user namespaces is widely supported by default on current HPC systems,  making \texttt{BEE} a highly usable framework. So, using Charliecloud we design \texttt{BEE-Charliecloud} for HPC platforms.

\begin{figure}[h]
    \centering
    \includegraphics[width=0.4\textwidth]{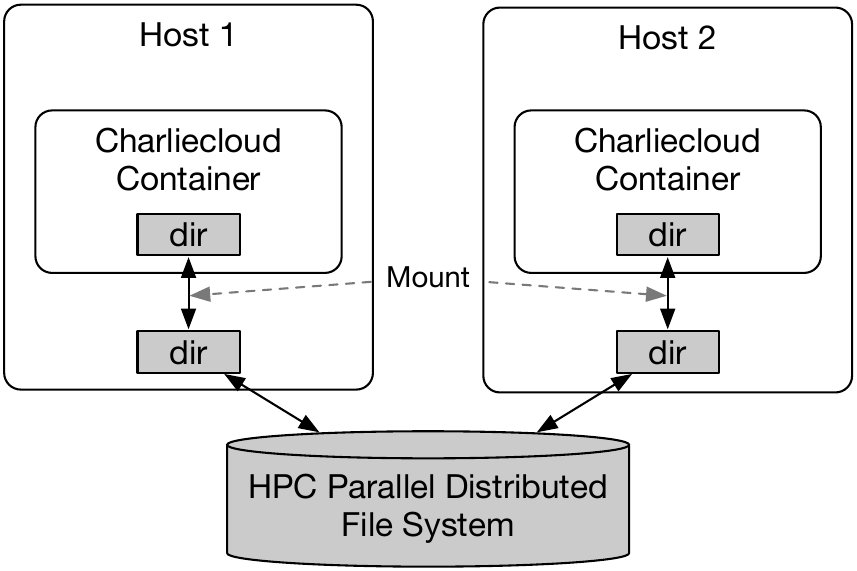}
    \caption{Shared Storage Design using Virtual IO}
    \label{bee-cc}
    \vspace*{-1em}
\end{figure}

When designing a \texttt{BEE backend}, one problem is identifying the suitable balance point between sharing and isolation of the runtime environment. Charliecloud offers comparable runtime isolation to Docker, that enables the user to pack all dependencies and tools in the image, avoiding additional application-specific configuration in the runtime environment. On the other hand, HPC applications usually require some degrees of data sharing: \textit{sharing via network} and \textit{sharing via storage}. 

\subsection{Network Design}
Sharing via network means processes running on different containers need to be able to share data via network. Charliecloud by default exposes all hardware network interfaces to its container runtime environment. Since HPC systems usually have interconnected networks between nodes, processes running in Charliecloud can communicate through the network interfaces on each node and get the benefits of any available network technology e.g., Infiniband. So, in \texttt{BEE-Charliecloud} we choose to the keep the default network settings. 

\subsection{Storage Design}
Sharing via storage is not natively supported by Charliecloud. By default each container only mounts the filesystem in the flattened input image. To enable sharing via storage, we use the \texttt{--bind} option  at container launch time to mount an user-specified host directory to a shared directory inside the container as mentioned in the user interface section. Since HPC systems usually have shared filesystem among nodes, processes can share data via the shared directory inside each container as shown in \textbf{Fig. \ref{bee-cc}} After each container is configured, \texttt{BEE-Charliecloud} automatically executes user's runscripts via the Charliecloud command \texttt{ch-run}. When initiating MPI parallel jobs, \texttt{BEE-Charliecloud} wraps \texttt{mpirun} command outside \texttt{ch-run} together with necessary MPI launching options. When deployed on HPC systems with Slurm resource manager, \texttt{BEE-Charliecloud} can interact with Slurm frontend to configure computing resources automatically. \textbf{Algorithm \ref{bee-cc-launch}} shows the launching logic of \texttt{BEE-Charliecloud}.

\begin{algorithm}
\caption{\texttt{BEE-Charliecloud} launching logic}
\label{bee-cc-launch}
\begin{algorithmic}[1]
\REQUIRE{Dockerized application (Docker image/Dockerfile)}
\REQUIRE{\texttt{BEE} configuration file (\texttt{beefile})}
\REQUIRE{Run scripts}
\STATE \texttt{pull/build\_docker(beefile)}
\STATE \texttt{flattened\_tar\_file $\leftarrow$ ch-docker2tar(docker\_image)}
\STATE \texttt{flattened\_filesystem $\leftarrow$ ch-tar2dir(flattened\_tar\_file)}
\STATE \texttt{compose\_ch-run\_options()}
\FOR{\textbf{each} \texttt{sequential run script} \textbf{in} \texttt{beefile} }
\STATE \texttt{slurm\_allocate\_resources()}
\STATE \texttt{ch-run(script)}
\ENDFOR
\FOR{\textbf{each} \texttt{parallel run script} \textbf{in} \texttt{beefile} }
\STATE \texttt{slurm\_allocate\_resources()}
\STATE \texttt{mpirun\_ch-run(mpi\_script)}
\ENDFOR

\end{algorithmic}
\end{algorithm}
\section{BEE-VM Design}
\label{bee-vm-section}

\begin{figure}[h]
    \centering
    \includegraphics[width=0.4\textwidth]{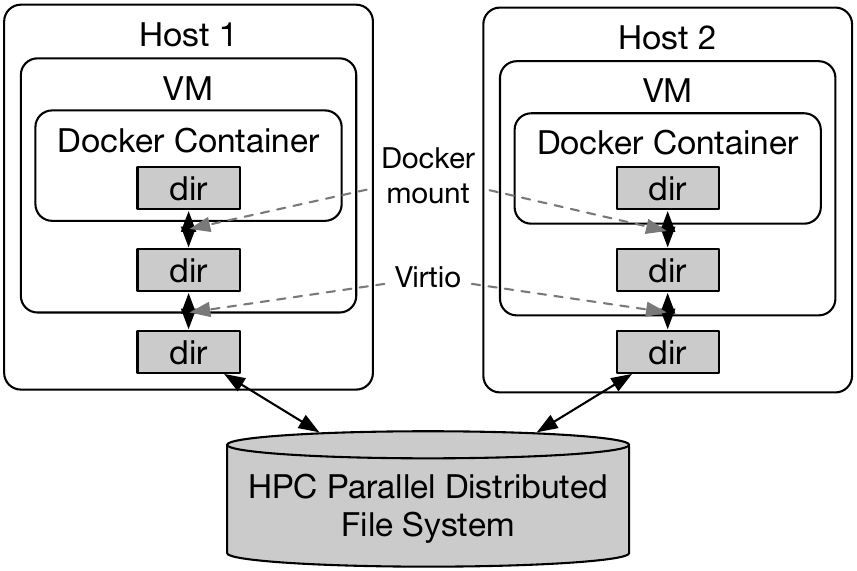}
    \caption{Shared Storage Design using Virtual IO}
    \label{bee-vm}
    \vspace*{-1em}
\end{figure}

To support \texttt{BEE} on older HPC systems that do not have Linux user namespace enabled, we design a second \texttt{BEE backend} for HPC system - \texttt{BEE-VM}, so that together with \texttt{BEE-Charliecloud} they make \texttt{BEE} support majority of the current HPC systems.
\texttt{BEE-VM} creates a VM layer on top of the host and then deploys Docker on the VM layer; as shown in \textbf{Fig. \ref{bee-backend}}. VM brings the isolation that enables us to run Docker even on a system with constraint. It utilizes Kernel-based Virtual Machine (KVM) (on by default in Linux), a hardware accelerated hypervisor, to provide bare-metal performance. As shown in \textbf{Fig. \ref{bee-vm}}, we run one VM per HPC host node and one Docker container per VM.


Similar to \texttt{BEE-Charliecloud}, besides running applications in isolated environment, another goal of the \texttt{BEE-VM} design is enabling two kinds of sharing, \textit{sharing via network} and \textit{sharing via storage}, between Docker container applications running on different machines. 
 
\subsection{Network Design}
The network design of \texttt{BEE-VM} mainly targets two functions: (1) \texttt{BEE} needs to remotely login and control each VM via SSH; (2) MPI needs network to share data between processes. 


In order to enable the SSH connection to the VM through the host, the hypervisor is configured to create dedicated virtual network interface card (vNIC) used for port forwarding that maps an unused port on the host to the SSH port on the VM. As for MPI, it cannot use the vNIC for SSH, since it usually uses a different (random) port for communication, so we cannot use port forwarding on a specific port. To handle this problem, we have two solutions. For a system with regular Ethernet, we create a second vNIC on each VM and connect all of them in a virtual subnet (i.e., multicast or P2P connection). We choose this over more straight forward approaches  (e.g., bridging) because this approach does not require any administrative privilege to the system. For a system with InfiniBand, we adopt Single Root Input/Output Virtualization (SR-IOV) to connect all VMs. To connect all the Docker containers together, we choose to start the Docker container in 'host network' in which all network configurations are exposed to the container, so that each container has the same connectivity as VM.

\subsection{Storage Design}
To share data between processes in Docker containers, we need to first build a shared filesystem between different VMs. Here, we use the Virtio feature \cite{russell2008virtio} in QEMU to map a host directory to a directory inside VMs. It only requires minimum configuration at VM boot time. Since HPC systems usually use shared filesystem (via NFS, Luster, etc.), each VM will also have the same file-sharing capability as long as they map to the same host directory. For data sharing in the Docker layer, we use the data volume mount feature in Docker to mount the shared folder inside VM to a directory in Docker. Since Docker runs as a process at the VM layer, mounting the data volume adds negligible overhead. 


\subsection{BEE-VM Deployment}
\textbf{Algorithm \ref{bee-launch}} shows the launching logic of \texttt{BEE-VM}. Before using \texttt{BEE-VM} the first time, users need to use the image builder provided by \texttt{BEE} to build the base VM image. The image is customized for \texttt{BEE-VM} that includes pre-installed softwares and settings. It only need to be done once. After that each VM will only work on a copy of the base image without any modification to the base image. During launch, \texttt{BEE-VM} first deploys VM on each HPC node (line 1 - 6). Next, the hostfile on each VM is setup in order to let each VM communicate with each other without using complicated IP addresses. The virtio storage is mounted in line 14. In the next stage, depending on what the user provides, \texttt{BEE-VM} will either pull the Docker image from public/private registries or build a new Docker image from a Dockerfile loaded into the local VM. Finally, \texttt{BEE} starts the application by launching from the first node (i.e., master node).

\begin{algorithm}
\caption{\texttt{BEE-VM} launching logic}
\label{bee-launch}
\begin{algorithmic}[1]
\REQUIRE{Pre-built QEMU Img. (only need to build once)}
\REQUIRE{Allocated host nodes: $H_1$, $H_2$,..., $H_k$}
\REQUIRE{Dockerized application (Docker image/Dockerfile)}
\REQUIRE{\texttt{BEE} configuration file (\texttt{beefile})}
\REQUIRE{Run scripts}

\FOR{i \textbf{in} 1 \textbf{to} \texttt{beefile}.num\_of\_nodes}
	\STATE \texttt{node\_i.copy\_img()}
	\STATE \texttt{node\_i.compose\_qemu\_arg(storage\_dir)}
	\STATE \texttt{node\_i.compose\_qemu\_arg(network\_conf)}
    \STATE \texttt{node\_i.start\_vm()}
\ENDFOR
\STATE \texttt{wait\_for\_all\_vm\_to\_become\_ready()}
\FOR{i in 1 \textbf{to} \texttt{beefile}.num\_of\_nodes}
\STATE \texttt{node\_i.set\_hostname()}
\FOR{j in 1 \textbf{to} \texttt{beefile}.num\_of\_nodes}
\STATE \texttt{node\_j.setup\_hostfile(node\_i.ip())}
\ENDFOR
\ENDFOR
\FOR{i in 1 \textbf{to} \texttt{beefile}.num\_of\_nodes}
\STATE \texttt{node\_i.mount\_via\_virtio()}
\ENDFOR
\FOR{i in 1 \textbf{to} \texttt{beefile}.num\_of\_nodes}
\STATE \texttt{node\_i.pull/build\_docker(beefile)}
\STATE 
\texttt{node\_i.conf\_docker\_storage(efs\_mnt)}
\STATE \texttt{node\_i.conf\_docker\_network(host\_mode)}
\STATE \texttt{node\_i.start\_docker('ssh daemon')}
\ENDFOR
\FOR{\textbf{each} \texttt{sequential run script} \textbf{in} \texttt{beefile} }
\STATE \texttt{node\_0.docker\_exec(script)}
\ENDFOR
\FOR{\textbf{each} \texttt{parallel run script} \textbf{in} \texttt{beefile} }
\STATE \texttt{node\_0.docker\_exec(mpi\_script)}
\ENDFOR

\end{algorithmic}
\end{algorithm}

\section{BEE-AWS Design}
\label{bee-aws-section}
AWS is one of the most widely used commercial cloud computing platforms. It offers great configuration flexibility towards computing software/hardware environment, network, storage, and security. Many researchers from both HPC and cloud community use AWS to run large scale applications. However, configuring the desired computing environment on AWS is tedious and can impede experiment workflow, especially for users with less knowledge or experience on AWS. Although many tools offer the ability to automate the computing environment setup process, they usually cannot create an environment suitable for HPC applications. Even though tools such as StarCluster \cite{starcluster}, do offer the capability of creating HPC-friendly environment on AWS, it suffers from two drawbacks. First, they do not offer the support of automatic Docker container deployment and execution. Users still need to manually deploy and run their applications. Second, they usually do not offer HPC/cloud cross-platform support. 

Here we design \texttt{BEE-AWS} as another \texttt{BEE backend}. \texttt{BEE-AWS} enables the same end-to-end automation on AWS as provided in \texttt{BEE-Charliecloud} and \texttt{BEE-VM} on HPC systems. Most importantly, \texttt{BEE-AWS} offers the same user interface and execution environment as \texttt{BEE-VM}, so the user only needs to make minimum modifications to the \texttt{BEE} configuration file (\texttt{beefile}) in order to switch to AWS. 

The computing resources are based on Xen-based VMs and they also known as EC2 instances. Since EC2 instances are VM-based, users are given full control inside each VM. So, here we configure them (via customized AMI) to enable Docker runtime. Users of \texttt{BEE} can specify desired instance type via \texttt{BEE} configuration file (\texttt{beefile}). Same as on the HPC platform, data sharing via network and storage also need to be handled.

\subsection{Network Design}
EC2 instances by default have network interconnect capability via the network in their infrastructures. However, they still need to be customized for HPC applications. First, as mentioned in \texttt{BEE-VM} design, MPI commonly uses random port for communication, so we need to create an EC2 security group that has a range of ports opened based on the MPI implementation specification. Second, for fast network interconnection, EC2 instances need to be placed in the same placement group. This makes sure the physical hardware allocated for each EC2 instance are close to each other so that the network is optimized for low latency and high throughput. As for network interconnection between Docker containers, we follow a similar choice made using \texttt{BEE-VM}, that enables 'host network' mode at launch time.

\subsection{Storage Design}
By default EC2 instances do not share filesystems. To enable file sharing, we choose to create Elastic File System (EFS) and mount EFS to each EC2 instance. This design has better performance than the master-slave based Network File System (NFS) adopted in Starcluster. We use the volume mounting feature of Docker to enable file sharing between Docker containers similar to \texttt{BEE-VM}.

\textbf{Algorithm \ref{bee-aws-launch}} shows that launching logic of \texttt{BEE-AWS}. We use Boto API to remotely launch EC2 instances on AWS (line 7 - 15). After that we use SSH connection to control each instance. 

\begin{algorithm}
\caption{\texttt{BEE-AWS} launching logic}
\label{bee-aws-launch}
\begin{algorithmic}[1]
\REQUIRE{Pre-built AMI (only need to build once)}
\REQUIRE{Dockerized application (Docker image/Dockerfile)}
\REQUIRE{\texttt{BEE} configuration file (\texttt{beefile})}
\REQUIRE{Run scripts}
\IF {\textit{user given EFS name} not exist}
\STATE \texttt{request\_creating\_efs(efs\_name)}
\WHILE{ \texttt{efs\_status(efs\_name)} != \texttt{active}}
\STATE \texttt{sleep()}
\ENDWHILE
\ENDIF

\STATE \texttt{initialize\_ec2\_service\_connection()}
\STATE \texttt{bee\_sg = create\_security\_group()}
\STATE \texttt{bee\_sg.authroize\_ingress('tcp', '22')}
\STATE \texttt{bee\_pg = create\_placement\_group()}
\FOR{i in 1 \textbf{to} \texttt{beefile}.num\_of\_nodes}
\STATE \texttt{bee\_ec2\_i = create\_ec2(bee\_sg, bee\_pg)}
\STATE \texttt{bee\_ec2\_i.start()}
\ENDFOR
\STATE \texttt{wait\_for\_all\_instance\_to\_become\_ready()}
\FOR{i in 1 \textbf{to} \texttt{beefile}.num\_of\_nodes}
\STATE \texttt{bee\_ec2\_i.set\_hostname()}
\FOR{j in 1 \textbf{to} \texttt{beefile}.num\_of\_nodes}
\STATE \texttt{bee\_ec2\_j.setup\_hostfile(bee\_ec2\_i.ip)}
\ENDFOR
\ENDFOR
\FOR{i in 1 \textbf{to} \texttt{beefile}.num\_of\_nodes}
\STATE \texttt{bee\_ec2\_i.create\_efs\_mount\_point()}
\STATE \texttt{bee\_ec2\_i.mount\_to\_efs(efs\_name)}
\STATE \texttt{bee\_ec2\_i.pull/build\_docker(beefile)}
\ENDFOR
\FOR{i in 1 \textbf{to} \texttt{beefile}.num\_of\_nodes}
\STATE \texttt{bee\_ec2\_i.pull/build\_docker(beefile)}
\STATE \texttt{bee\_ec2\_i.conf\_docker\_storage(efs\_mnt)}
\STATE \texttt{bee\_ec2\_i.conf\_docker\_network(host\_mode)}
\STATE \texttt{bee\_ec2\_i.start\_docker('ssh daemon')}
\ENDFOR
\FOR{\textbf{each} \texttt{sequential run script} \textbf{in} \texttt{beefile} }
\STATE \texttt{bee\_ec2\_0.docker\_exec(script)}
\ENDFOR
\FOR{\textbf{each} \texttt{parallel run script} \textbf{in} \texttt{beefile} }
\STATE \texttt{bee\_ec2\_0.docker\_exec(mpi\_script)}
\ENDFOR
\end{algorithmic}
\end{algorithm}

\section{BEE-OpenStack Design}
\label{bee-openstack-section}
OpenStack is a cloud operating system that is able to manage large pools of computing, storage, and network resources. It has been widely deployed in both research facilities and cloud computing environments. To bring the same unified execution environment and end-to-end automation to OpenStack, we build \texttt{BEE-OpenStack} as another \texttt{BEE backend}.

Unlike AWS, the computing sources of OpenStack can be either bare-metal machines or VM. In either case, users are given full control inside the operating system. So, similar to \texttt{BEE-AWS} we enable Docker runtime inside each OpenStack instance. 

\subsection{Network Design}
On our OpenStack test environment (\texttt{Chameleon Cloud}), network interconnects are enabled by default between instances. So, we do not need to further configure it. For OpenStack infrastructures that need customized network, \texttt{BEE} will configure it automatically to ensure network interconnect capabilities between instances. The detail is omitted here. Similar to before, we use 'host network' mode for Docker containers.

\subsection{Storage Design}
Similar to \texttt{AWS}, by default, OpenStack instances do not share filesystems. On our OpenStack test environment (\texttt{Chameleon Cloud}), there is no OpenStack-managed storage system. So, we adopt NFS based file sharing between master instance (first instance) and worker instances. We use the volume mounting feature of Docker to enable file sharing between Docker containers similar to \texttt{BEE-VM} and \texttt{BEE-AWS}

\textbf{Algorithm \ref{bee-openstack}} shows the launching logic of \texttt{BEE-OpenStack}. Here we first use OpenStack CLI client to launch a pre-built Stack template for BEE, and then use SSH to control each instance.

\begin{algorithm}
\caption{\texttt{BEE-OpenStack} launching logic}
\label{bee-openstack}
\begin{algorithmic}[1]
\REQUIRE{Pre-built OpenStack Img. (only need to build once)}
\REQUIRE{Dockerized application (Docker image/Dockerfile)}
\REQUIRE{\texttt{BEE} configuration file (\texttt{beefile})}
\REQUIRE{Run scripts}
\STATE \texttt{initialize\_nova\_service\_connection()}
\STATE \texttt{create\_new\_sshkey()}
\STATE \texttt{launch\_bee\_stack(\texttt{beefile})}
\STATE \texttt{wait\_for\_all\_instance\_to\_become\_ready()}

\FOR{i in 1 \textbf{to} \texttt{beefile}.num\_of\_nodes}
\STATE \texttt{bee\_os\_i.set\_hostname()}
\FOR{j in 1 \textbf{to} \texttt{beefile}.num\_of\_nodes}
\STATE \texttt{bee\_os\_j.setup\_hostfile(bee\_os\_i.ip)}
\ENDFOR
\ENDFOR
\STATE \texttt{bee\_os\_0.create\_nfs\_mount\_point()}
\FOR{i in 1 \textbf{to} \texttt{beefile}.num\_of\_nodes}
\STATE \texttt{bee\_os\_i.mount\_to\_nfs(bee\_os\_0.ip())}
\ENDFOR
\FOR{i in 1 \textbf{to} \texttt{beefile}.num\_of\_nodes}
\STATE \texttt{bee\_os\_i.pull/build\_docker(beefile)}
\ENDFOR
\FOR{i in 1 \textbf{to} \texttt{beefile}.num\_of\_nodes}
\STATE \texttt{bee\_os\_i.pull/build\_docker(beefile)}
\STATE \texttt{bee\_os\_i.conf\_docker\_storage(nfs\_mnt)}
\STATE \texttt{bee\_os\_i.conf\_docker\_network(host\_mode)}
\STATE \texttt{bee\_os\_i.start\_docker('ssh daemon')}
\ENDFOR
\FOR{\textbf{each} \texttt{sequential run script} \textbf{in} \texttt{beefile} }
\STATE \texttt{bee\_os\_0.docker\_exec(script)}
\ENDFOR
\FOR{\textbf{each} \texttt{parallel run script} \textbf{in} \texttt{beefile} }
\STATE \texttt{bee\_os\_0.docker\_exec(mpi\_script)}
\ENDFOR
\end{algorithmic}
\end{algorithm}
\section{Evaluation}
\label{evaluation-section}
We evaluate the four \texttt{BEE backends} on four different platforms: For \texttt{BEE-Charliecloud}
and \texttt{BEE-VM}, we test them on the bare-metal envionment on \texttt{Chameleon Cloud} at Texas Advanced Computing Center (TACC). Each node is equipped with two Intel Xeon E5-2670 v3 CPU (clock frequency at 2.30 GHz) with 128 GB DRAM. Each node is connected with Mellanox ConnectX-3 Infiniband card with peak transfer speed at 10 Gbps.  For \texttt{BEE-AWS}, we choose to use \texttt{c3.4xlarge} EC2 type at AWS Oregon region. Each node is equipped with Intel Xeon E5-2680 v2 CPUs and 30GB DRAM. On \texttt{BEE-OpenStack}, we choose the OpenStack environment at \texttt{Chameleon Cloud}@University of Chicago. We focus on evaluating three kinds of performance that are most important for HPC applications: \textit{computation}, \textit{storage}, and \textit{network}. For each test, the comparison baseline is the native performance provided on each platform without using \texttt{BEE} or any additional encapsulated runtime system. Note: all platforms, except AWS, provide access to the bare-metal hardware, so baseline performance is the native performance on the hardware. For AWS, its Xen-based VM is an inseparable part of the platform and the underlying physical hardware is inaccessible to general users, so its baseline is the performance inside VMs.

\begin{figure*}[t]
    \centering
    \begin{subfigure}[t]{0.49\textwidth}
        \includegraphics[width=\textwidth]{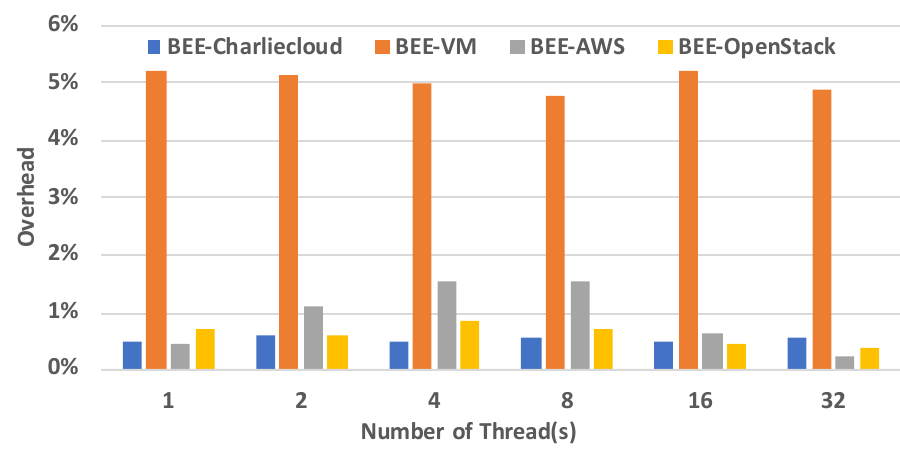}
        \caption{Compute intensive workload (BT)}
    \end{subfigure}
    \begin{subfigure}[t]{0.49\textwidth}
        \includegraphics[width=\textwidth]{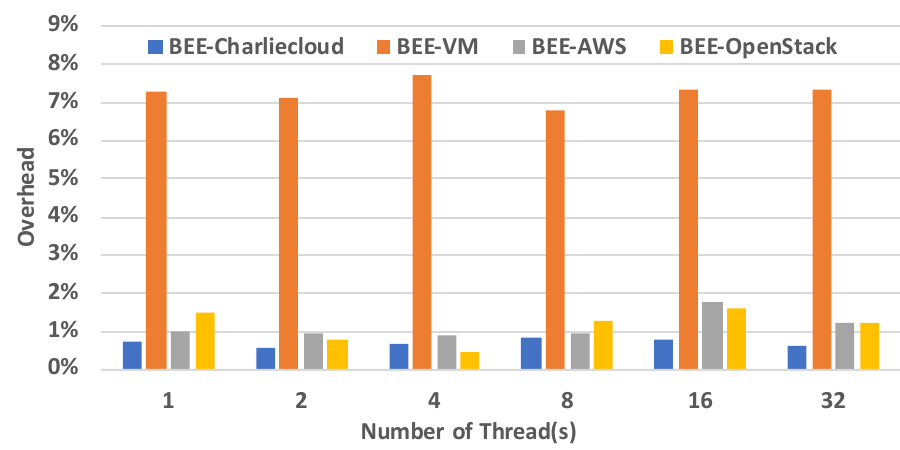}
        \caption{Memory intensive workload (IS)}
    \end{subfigure}
    \vspace*{-0.5em}
    \caption{Performance overhead compare with native performance provided on each corresponding platform.}
    \label{comp}
    \vspace*{-1em}
\end{figure*}

\begin{figure*}[t]
    \centering
    \begin{subfigure}[t]{0.49\textwidth}
        \includegraphics[width=\textwidth]{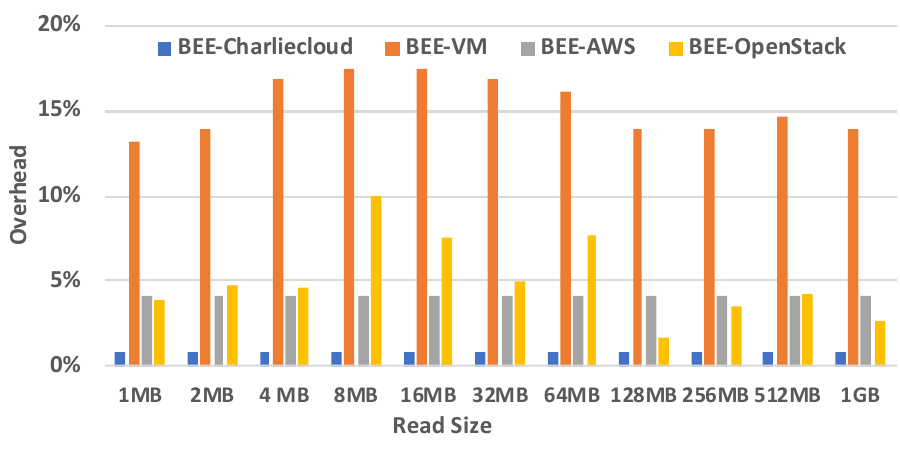}
    \end{subfigure}
    \begin{subfigure}[t]{0.49\textwidth}
        \includegraphics[width=\textwidth]{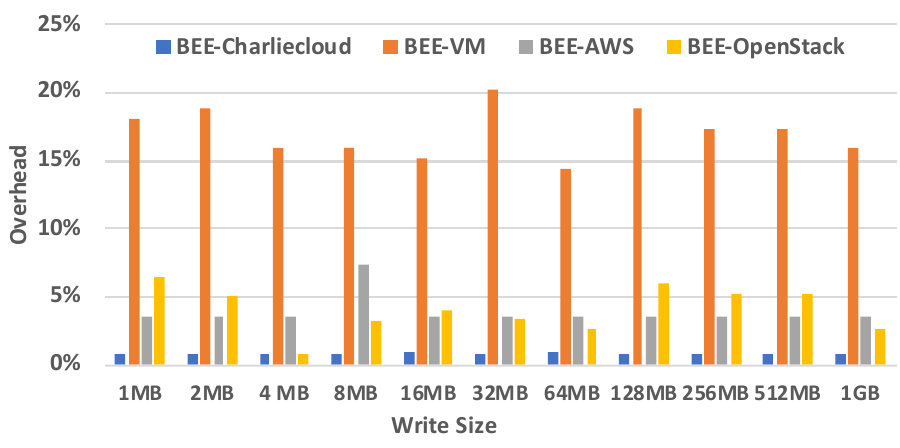}
    \end{subfigure}
    \vspace*{-0.5em}
    \caption{Storage I/O read/write speed overhead compare with native speed provided on each corresponding platform.}
    \label{io}
    \vspace*{-2em}
\end{figure*}

\subsection{Computational Performance}
Computational performance of a platform is one of the most important capabilities for HPC applications. In this section, we compare the computational performance of all four \texttt{BEE backends} with the baseline. We choose one compute intensive benchmark test and one memory intensive benchmark test from the OpenMP version NAS Parallel Benchmarks (NPB)\cite{npb} in our evaluation. We test each benchmark running one to 32 threads (cores) to further show the computational performance on multi-thread environment. 
\subsubsection{Compute intensive workload}
For compute intensive workload, we choose Block Tri-diagonal solver (BT) benchmark test with input matrix size of $102^3$ (problem size: \texttt{class B}).

As we can see in \textbf{Fig. \ref{comp}(a)}, for compute intensive workload, all four \texttt{BEE backends} have low performance overhead: \texttt{BEE-Charliecloud} 0.4\% - 0.6\% (avg. 0.5\%); \texttt{BEE-VM} 4.8\% - 5.2\% (avg. 5.0\%); \texttt{BEE-AWS} 0.2\% - 1.6\% (avg. 0.8\%); \texttt{BEE-OpenStack} 0.3\% - 0.9\% (avg. 0.6\%).
    
\subsubsection{Memory intensive workload}
For memory intensive workload, we choose Integer Sort  (IS) test suit with 134217728 input integer (problem size : \texttt{class C}). 

As we can see in \textbf{Fig. \ref{comp}(b)}, for memory intensive workload all four \texttt{BEE backends} also have low performance overhead: \texttt{BEE-Charliecloud} 0.6\% - 0.9\% (avg. 0.7\%); \texttt{BEE-VM} 6.9\% - 7.7\% (avg. 7.1\%); \texttt{BEE-AWS} 0.9\% - 1.8\% (avg. 1.1\%); \texttt{BEE-OpenStack} 0.5\% - 1.7\% (avg. 1.0\%). In addition, both kinds of workload also exhibit similar speedup comparing with their baseline counterparts when we increase the number of threads.

 \begin{figure*}[t]
    \centering
    \begin{subfigure}[t]{0.245\textwidth}
        \includegraphics[width=\textwidth]{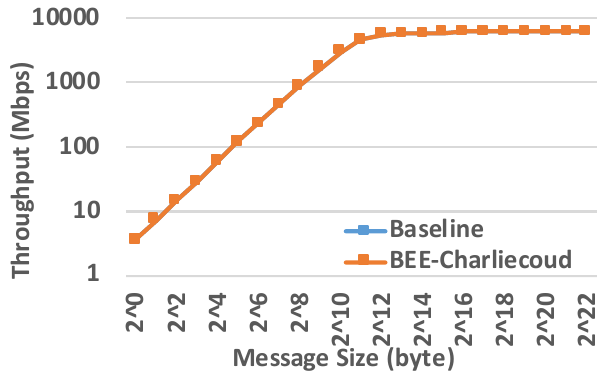}
    \end{subfigure}
    \begin{subfigure}[t]{0.245\textwidth}
        \includegraphics[width=\textwidth]{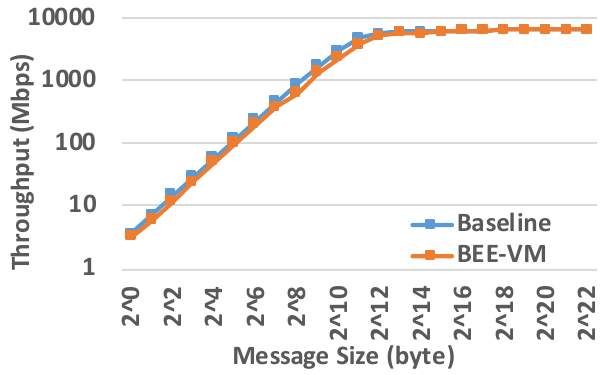}
    \end{subfigure}
    \begin{subfigure}[t]{0.245\textwidth}
        \includegraphics[width=\textwidth]{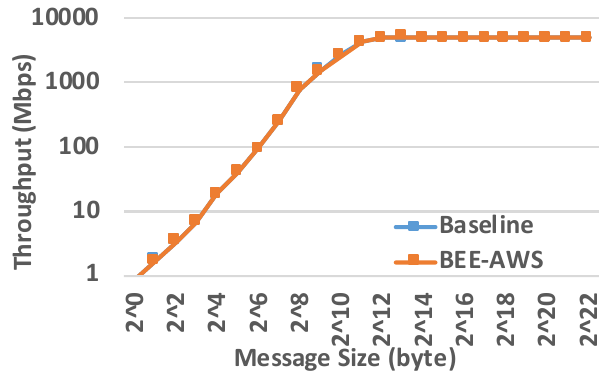}
    \end{subfigure}
    \begin{subfigure}[t]{0.245\textwidth}
        \includegraphics[width=\textwidth]{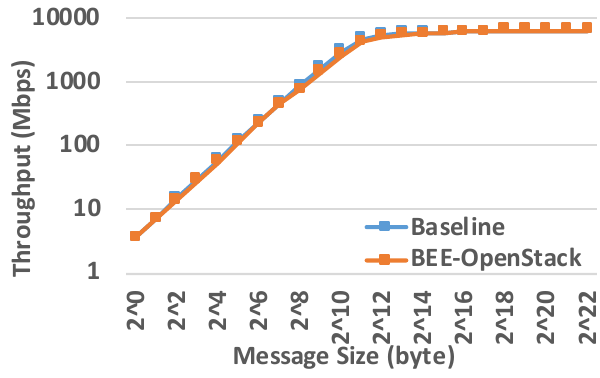}
    \end{subfigure}
    \caption{P2P Network Throughput Comparison}
    \label{net-band}
\end{figure*}
\begin{figure*}[t]
    \centering
    \begin{subfigure}[t]{0.245\textwidth}
        \includegraphics[width=\textwidth]{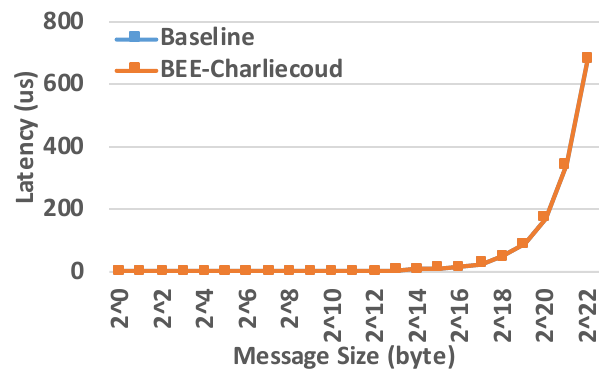}
    \end{subfigure}
    \begin{subfigure}[t]{0.245\textwidth}
        \includegraphics[width=\textwidth]{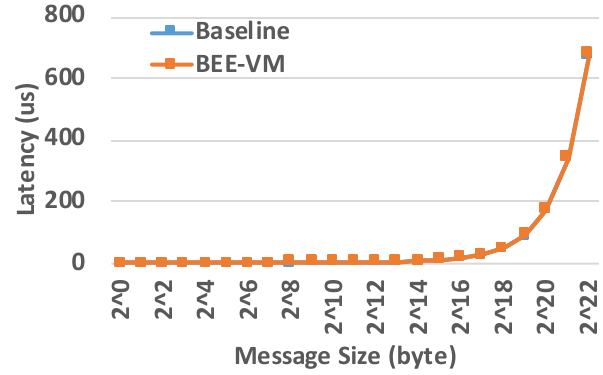}
    \end{subfigure}
    \begin{subfigure}[t]{0.245\textwidth}
        \includegraphics[width=\textwidth]{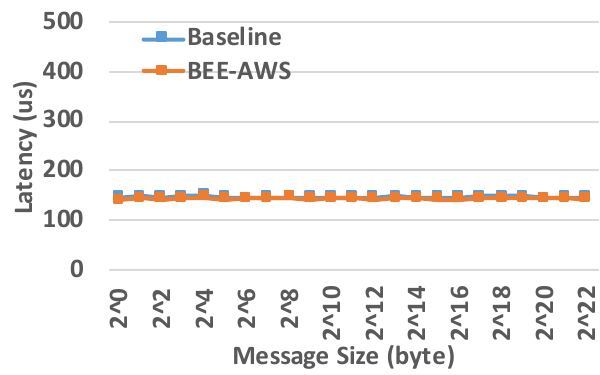}
    \end{subfigure}
    \begin{subfigure}[t]{0.245\textwidth}
        \includegraphics[width=\textwidth]{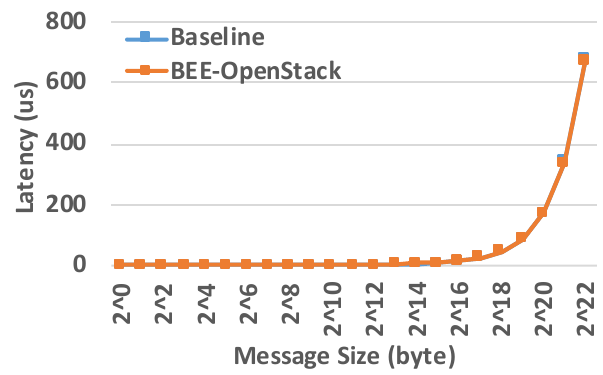}
    \end{subfigure}
    \caption{P2P Network Latency Comparison}
    \label{net-lat}
\end{figure*}

\begin{figure*}[t]
    \centering
    \begin{subfigure}[t]{0.245\textwidth}
        \includegraphics[width=\textwidth]{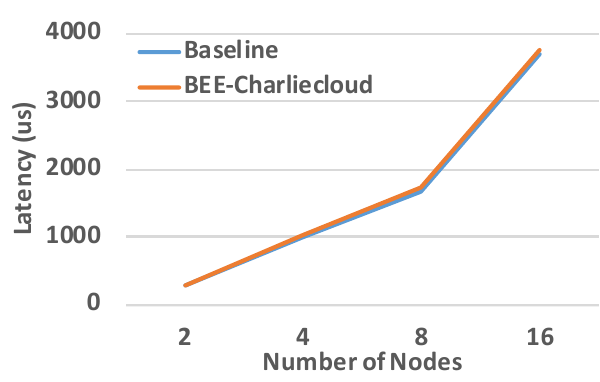}
    \end{subfigure}
    \begin{subfigure}[t]{0.245\textwidth}
        \includegraphics[width=\textwidth]{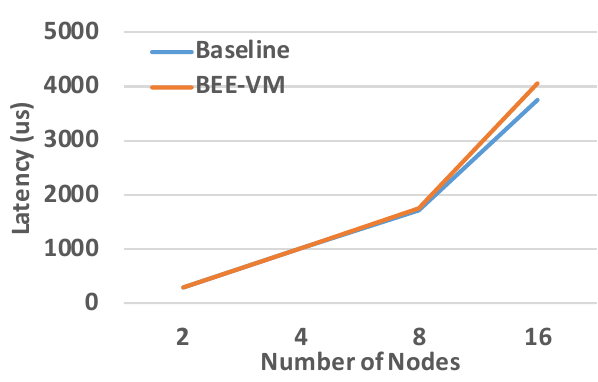}
    \end{subfigure}
    \begin{subfigure}[t]{0.245\textwidth}
        \includegraphics[width=\textwidth]{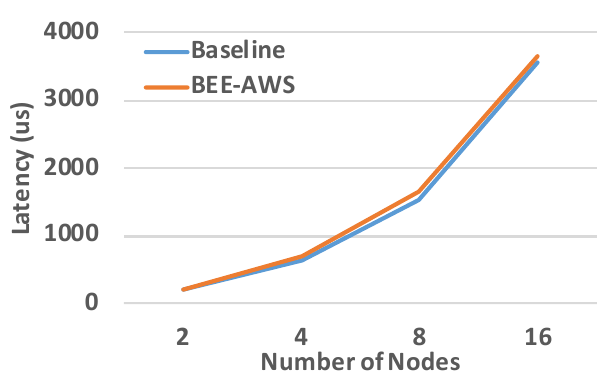}
    \end{subfigure}
    \begin{subfigure}[t]{0.245\textwidth}
        \includegraphics[width=\textwidth]{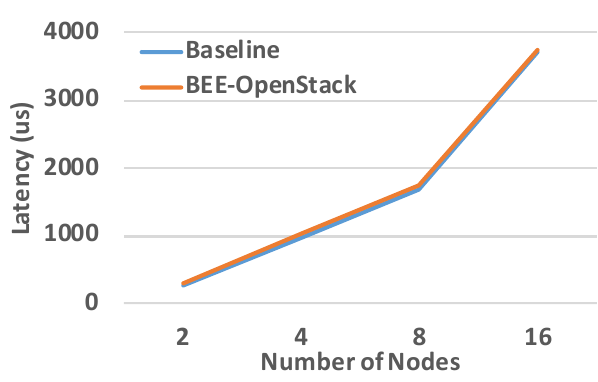}
    \end{subfigure}
    \caption{All-to-all Network Latency Comparison}
    \label{net-lat-all}
    \vspace*{-2em}
\end{figure*}
\subsection{Storage I/O }
Storage I/O is another key component for many HPC applications. For example, large-scale simulations often need to first load large datasets before computation and frequently dump checkpoints during computation and once the simulation has concluded. Later, these results may be used for other simulations or for analytics. I/O performance plays an important role for the overall performance of HPC applications. 

To evaluate the storage I/O performance of the four \texttt{BEE backends} and compare with the baseline performance, we use the Linux built-in command -- \texttt{dd}. Specifically, to benchmark write performance, we use the \texttt{dd} command to write a file with data from \texttt{/dev/zero}. As for read performance, we use the \texttt{dd} command to read out the file just saved and write to \texttt{/dev/null}. To avoid reading from cache, we use \texttt{echo 3 > /proc/sys/vm/drop\_caches} command to force the system to clear all cached data between each write and read. Usually this file is read-only inside a container, so we issue the command outside the container, achieving the same cache flushing effect. The file used for write and read is placed in the directory that is shared between instances or containers. We test different file sizes ranging from 1 MB to 1 GB. To eliminate noise and variation, we repeat each test 1000 times. \textbf{Fig. \ref{io}} shows the read and write performance on the four \texttt{BEE backends} comparing with the baseline (i.e., the IO performance on each platform without using \texttt{BEE}). As we can see, \texttt{BEE-Charliecloud} produces negligible read and write overhead (~0.08\%). \texttt{BEE-VM} uses both VM and Docker, so it produces relative higher overhead, 13.1\% - 17.5\% (avg. 15.2\%) for read and 15.1\% - 20.1\% (avg. 17.1\%) for write. \texttt{BEE-AWS} produces steady 0.4\% overhead for read and 0.3\% overhead for write. \texttt{BEE-OpenStack} produces 1.7\% - 7.6\% (avg. 5.0\%) for read overhead and 2.7\% - 6.4\% (avg. 4.1\%) for write overhead.

\subsection{Network}
Finally, we evaluate the network performance of the four \texttt{BEE backends}. We use the HPCBench \cite{hpcbench} to measure the bandwidth and latency when transferring data of different sizes between two processes in containers or on the platform without using \texttt{BEE}. \texttt{BEE-VM} and \texttt{BEE-OpenStack} use the Infiniband (IB) on \texttt{Chameleon Cloud} via SR-IOV. \texttt{BEE-Charliecloud} and \texttt{BEE-AWS} use the Ethernet connection. \textbf{Fig. \ref{net-band}} and \textbf{Fig. \ref{net-lat}} show the point-to-point (P2P) network performance. MPI communication function calls between two nodes (one process per node) were used to test the average throughput and latency when transferring different message sizes. It can be seen that all four \texttt{BEE backends} can provide similar network bandwidth and latency compared to the baseline. \textbf{Fig. \ref{net-lat-all}} also show the latencies of all-to-all collective communication in MPI. It shows that all four \texttt{BEE backends} still provide similar network performance large scale.

\section{Vector Particle-In-Cell Case Study}
\label{case-study-section}
 In this section, we showcase a widely used HPC application, Vector Particle-In-Cell (VPIC) plasma physics code \cite{bowers20080, bowers2008ultrahigh, bowers2009advances}, on the four \texttt{BEE backends}. VPIC is a general purpose particle-in-cell plasma simulation code for modeling kinetic plasmas in multiple spatial dimensions. VPIC is a memory bound application that runs on multiple nodes using MPI and pthreads. It has been optimized for modern computing architectures by using short-vector, single-instruction-multiple-data (SIMD) instructions and cache optimization. Before the simulation begins, VPIC needs to load an input deck and user configuration files. When computation is finished, VPIC writes the output. With flexible checkpoint-restart semantics,  VPIC allows checkpoint files to be read as input for subsequent simulations. Moreover, VPIC has a native I/O format that interfaces with the high-performance visualization software Ensight and Paraview.

\begin{figure}[h]
	\vspace*{-1em}
    \centering
    \includegraphics[width=0.45\textwidth]{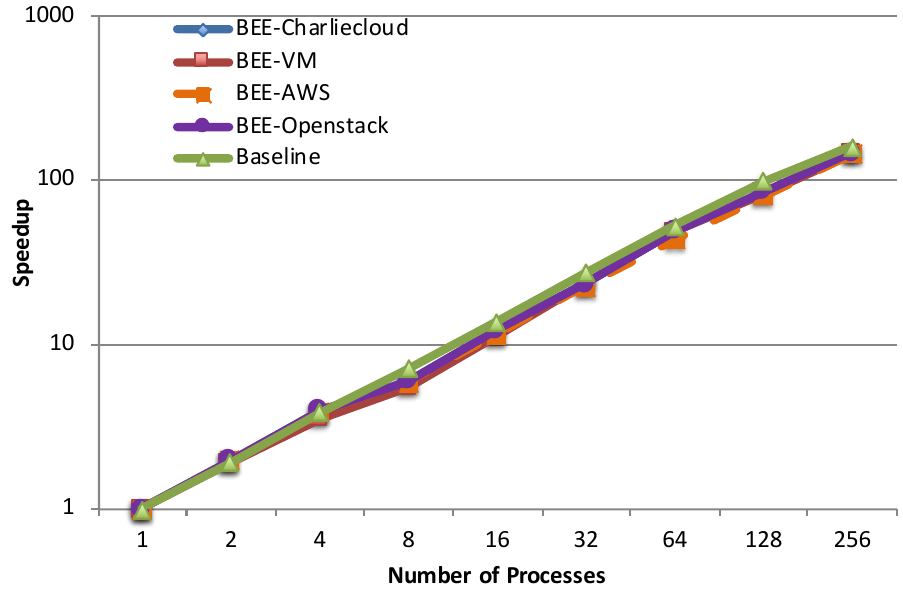}
    \caption{VPIC scale up test}
    \label{vpic-test}
    \vspace*{-1em}
\end{figure}

We test VPIC on different \texttt{BEE backends} using 1 to 256 processes. As shown in \textbf{Fig. \ref{vpic-test}}, all four \texttt{BEE backends} exhibit similar speedup compared with the baseline i.e., the speedup we can get on the bare-metal environment on \texttt{Chameleon Cloud} at Texas Advanced Computing Center (TACC).

\section{Related Work}
\label{related-work-section}
\subsection{StarCluster}
StarCluster \cite{starcluste} is an open source cluster launching tool for AWS that aims to automation the process of creating a HPC cluster-like computing environment on AWS. It uses a configuration file to let users specify desired configuration of the cluster e.g., number of instances. Similar to \texttt{BEE-AWS} it uses BOTO APT \cite{BOTOAPI} to communicate to AWS regards creating and launching EC2 instances and other resources. Although it offers automation, it suffers from three major drawbacks. First, it does not offer automatic deployment of Dockerized application. Current Docker users still need to manually copy Dockerized application to EC2 instances, configure the network and storage of each Docker container, then launch the application. Second, even not using Docker, users still need to manually login to the EC2 instances to deploy and launch their applications. Finally, it does not have HPC/cloud cross-platform portability.

\subsection{Amazon Elastic Container Service}
Amazon Elastic Container Service (ECS) \cite{awscontainer} offers the capability of automating creating a cluster of EC2 instances on AWS and then deploying the Docker container on each instance. However, it does not create a shared file system automatically; users still need to manually configure EFS or S3 storage. In addition, it only targets AWS platform with no  portability to other platforms.

\subsection{Kubernetes}
Kubernetes \cite{kubernetes} is widely used open-source system for automating deployment of containerized applications. It supports many mainstream cloud platforms e.g., AWS, Google Compute Engine, Azure, IMB Cloud, etc. However, as discussed in \cite{kubernetes-challenges}, Kubernetes is designed for service-based workload on those platforms, so they are not suitable HPC applications. In addition, it has limited support on HPC systems.


\section{Conclusions}
\label{conclusions-section}
In this work, we first discussed several workflow and productivity problems that current HPC users are facing. Then, we analyzed the potential of integrating commonly used Linux container technology into the existing HPC systems to build a consistent container environment for HPC users. Following this goal, we proposed a container-enabled framework, \texttt{BEE}, and four \texttt{BEE backends} for HPC and cloud systems that allow users to use the compute resources of both HPC systems and cloud computing systems with little extra configuration. Experiments show that \texttt{BEE} can achieve comparable performance. Finally, a case study on a widely used VPIC simulation tool was tested on \texttt{BEE} for performance comparison. 

\section{Acknowledgement}
This work was funded by the US Government contract DE-AC52-06NA25396 for Los Alamos National Laboratory, operated by Los Alamos National Security,
LLC, for the US Department of Energy. This work was also supported by NSF Award No. 1513201. Results presented in this paper were obtained using the Chameleon Cloud sponsored by the National Science Foundation. The publication has been assigned the LANL identifier LA-UR-18-27912.

\bibliographystyle{IEEEtran}
\bibliography{refer} 

\end{document}